\documentclass[aps,pra,notitlepage,superscriptaddress,showpacs]{revtex4-2}
\usepackage[LGR,T1]{fontenc}
\usepackage{color}
\usepackage{amsmath}
\usepackage{amssymb}
\usepackage{graphicx}
\usepackage{esint}
\usepackage[normalem]{ulem}

\makeatletter

\DeclareRobustCommand{\greektext}{%
  \fontencoding{LGR}\selectfont\def\encodingdefault{LGR}}
\DeclareRobustCommand{\textgreek}[1]{\leavevmode{\greektext #1}}
\ProvideTextCommand{\~}{LGR}[1]{\char126#1}

\makeatother

\begin{document}

\title{Coherent phase control of orbital-angular-momentum light-induced torque in a double-tripod atom-light coupling scheme}

\author{Hamid R. Hamedi}
\email{hamid.hamedi@tfai.vu.lt}
\affiliation{Institute of Theoretical Physics and Astronomy, Vilnius University,
LT-10257 Vilnius, Lithuania}

\author{Viačeslav Kudriašov}
\email{viaceslav.kudriasov@ff.vu.lt }
\affiliation{Institute of Theoretical Physics and Astronomy, Vilnius University,
LT-10257 Vilnius, Lithuania}

\author{Mažena Mackoit Sinkevičienė}
\email{mazena.mackoit-sinkeviciene@ff.vu.lt }
\affiliation{Institute of Theoretical Physics and Astronomy, Vilnius University,
LT-10257 Vilnius, Lithuania}

\author{Julius Ruseckas}
\email{julius.ruseckas@gmail.com }
\affiliation{Baltic Institute of Advanced Technology, LT-01403 Vilnius, Lithuania}

\begin{abstract}
We investigate a phase-controllable mechanism for generating optical
torque in a five-level double-tripod (DT) atom-light coupling scheme
interacting with four strong coherent control fields as well as two
weak optical vortex probe beams carrying orbital angular momentum
(OAM). The spatial phase gradients of the OAM-carrying probes induce
a quantized torque that is transferred to the atoms, rotating them
and generating a directed atomic flow within an annular geometry.
Analytical solutions of the optical Bloch equations under steady-state
conditions show that the induced torque and resulting rotational motion
exhibit high sensitivity to phase variations. We show that the DT
system coherently reconfigures into either coupled $\Lambda$
or double-$\Lambda$ schemes depending on the relative phases,
with each configuration exhibiting distinct quantized torque characteristics.
This enables precise phase control of the atomic current flow, with potential applications in quantum control, precision measurement, and quantum information processing.
\end{abstract}

\maketitle

\section{Introduction}

\textcolor{black}{The interaction between coherent light and atomic
systems leads to a wide range of intriguing quantum phenomena, driven
by atomic coherences and light-induced control of medium properties
\citep{Walls1994,Scully1997,Meystre2001,Anglin2002}. One of the most
extensively studied effects in this domain is electromagnetically
induced transparency (EIT) \citep{Harris1997,Lukin2003,Fleischhauer2005},
a quantum interference phenomenon that allows an otherwise opaque
medium to become transparent under specific conditions. EIT plays
a crucial role in slow light experiments, where the group velocity
of light is significantly reduced due to the presence of atomic coherence
\citep{Zhang2008,Juzeliunas2002,Juzeliunas2004,Meng2014,Siverns2019,Lukin2000}.
Slow light and EIT have been studied in three- and four level systems,
particularly in lambda and tripod configurations \citep{Juzeliunas2002,Juzeliunas2004,Fleischhauer2000,Hamedi2017},
each employing distinct pathways for coherence-based control of light
propagation. Moving beyond these schemes, the extension of EIT to
more complex multilevel atomic systems \citep{Paspalakis2002} has
opened new avenues for enriched quantum dynamics and enhanced light-matter
interaction control. Among these, the double tripod (DT) system  \citep{Meng2014,Bao2011}
has gained particular attention due to its unique coherence properties
and relevance to nonlinear optics and quantum information processing.
}Consisting of two upper and three lower atomic levels, coupled by
six optical fields in total, DT represents a highly controllable quantum
medium. A particularly interesting effect in the DT system is spinor
slow light (SSL) \textcolor{black}{\citep{Meng2014},}which introduces
additional internal degrees of freedom for controlling light-matter
interactions and has implications for structured light dynamics in
multilevel atomic ensembles.

In parallel with EIT and slow light, a fundamentally important\textcolor{black}{{}
concept in modern quantum optics is the orbital angular momentum (OAM)
of light \citep{Allen92,Allen99}. OAM-carrying vortex beams possess
a helical wavefront and can impart quantized angular momentum to atoms,
leading to novel interactions \citep{Yu2021}. A demonstration by
Radwell }\textit{\textcolor{black}{et al}}\textcolor{black}{. \citep{Radwell2015}
showed that the coherent coupling between OAM-carrying light and atomic
ensembles can lead to spatially dependent EIT. This was later extended
theoretically to more complex atomic schemes, where the interaction
of vortex beams with combined tripod and \textgreek{L}-type configurations
enabled the coherent transfer of OAM to a probe field in a resonant
medium \citep{HamidOE}. The influence of OAM on atomic systems has
been investigated in various contexts, including azimuthally dependent
EIT \citep{Radwell2015,HamidOE,Tarak2017}, direct OAM transfer between light and
atomic states \citep{Akamatsu2003,Ruseckas2013,Hamedi2018transfer,Hamedi2020transfer}, magnetometry \citep{Castellucci21},
and mechanical effects induced by structured light \citep{Babiker1994,Andersen2006,Lembessis2010}.
These interactions of OAM with atomic ensembles introduce additional
degrees of freedom to light-matter coupling, offering enhanced control
over atomic coherence and state manipulation, and opening pathways
for novel effects applicable to quantum computation, quantum information
storage and quantum teleportation \citep{Alison2011,Liu2020}. This
flexibility enables enhanced manipulation of quantum states, tailored
optical responses, and new coherence phenomena, all essential for
understanding the complex behavior of structured light fields in multi-level
atomic systems.}
 
\textcolor{black}{A particularly fascinating effect of OAM-carrying
light is the generation of optical torque in atomic media, where OAM light
fields transfer angular momentum to atoms, inducing mechanical rotation
or atomic motion \citep{Miles2011,Lembessis2010,Andersen2006,Babiker1994}.
A previous study investigated torque effects in a three-level $\Lambda$ system driven by two OAM beams, revealing complex exchanges of angular momentum between light and matter \citep{Lembessis2010}. In that case, the magnitude of the induced torque was determined solely by the beam amplitudes and the detuning of the fields. In this work, we propose phase control of the optical torque as a new degree of freedom in a more intricate five-level DT scheme. The DT system interacts with four strong control fields and two weak OAM-carrying probe fields, where the multiple excitation pathways make it an effective platform for exploring the mechanical impact of structured light. Importantly, the strong control fields form a closed loop, and their relative phase can directly influence the optical response of the probe fields. This feature is absent in a double-$\Lambda$ configuration \citep{Hamedi2018transfer}, where OAM probe beams are part of the closed loop. Because the intensity of OAM beams vanishes at the vortex core, the loop is effectively broken at those points, and no well-defined relative phase can be established—rendering phase control of the induced torque impossible in that setting. In contrast, in the DT scheme the OAM probe beams that generate torque lie outside the closed loop formed entirely by non-vortex control fields. As a result, the loop remains intact everywhere, independent of the probe’s vortex structure, making it possible to exert phase control over the probe properties, including the induced optical torque. We further show that by adjusting the relative phases of the control fields, the DT system can be coherently reconfigured into either coupled $\Lambda$ subsystems or an effective double-$\Lambda$ configuration. This reconfigurability enables tunable optical torque and controlled rotational motion of atoms, offering a mechanism for precise manipulation of atomic currents in annular geometries. Understanding the role of torque in this regime opens new avenues for controlling atomic motion and engineering advanced quantum optical systems.}

\section{Theoretical model}

\subsection{The system}

\begin{figure}
\includegraphics[width=0.5\columnwidth]{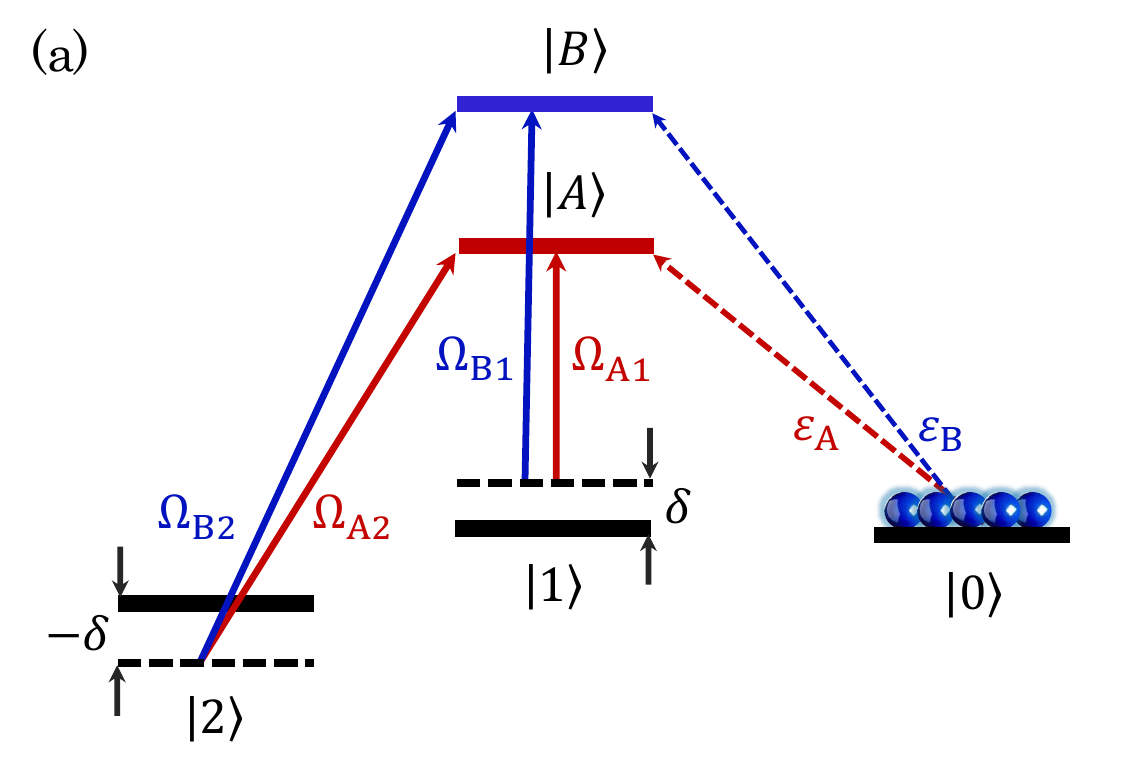}\includegraphics[width=0.5\columnwidth]{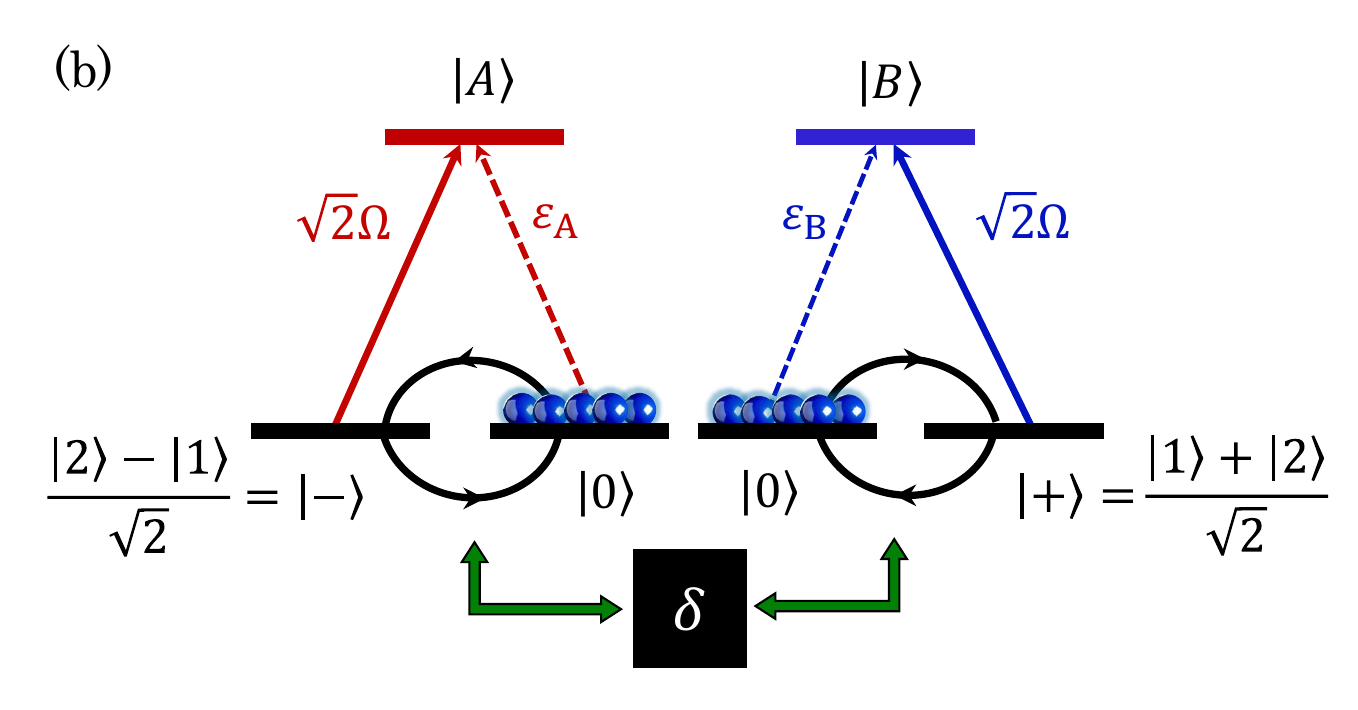}

~~~~~~~~~~~~~~~~~~~~~~\includegraphics[width=0.45\columnwidth]{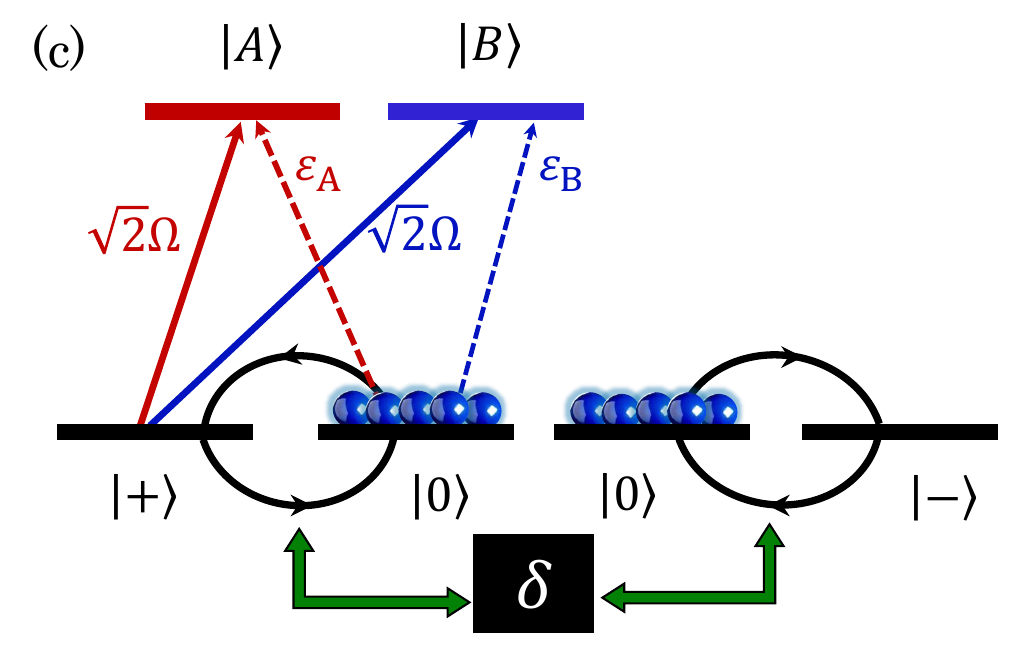}

\caption{\label{fig:1} (a) Schematic of the double-tripod (DT) configuration,
in which three ground states $\ensuremath{|0\rangle}$, $|1\rangle$,
$\ensuremath{|2\rangle}$ are coupled to two excited states $\ensuremath{|A\rangle}$,
$\ensuremath{|B\rangle}$ via two probe fields $\ensuremath{\varepsilon_{A}}$
and $\ensuremath{\varepsilon_{B}}$, along with four strong control
fields $\ensuremath{\Omega_{A1}}$, $\Omega_{A2}$, $\Omega_{B1}$,
$\ensuremath{\Omega_{B2}}$. (b) Representation of two coupled $\Lambda$-systems
with a common effective Rabi frequency $\sqrt{2}\Omega$. This model
is equivalent to the DT system, when all four coupling fields have
equal amplitude $\Omega$, the relative phase is $\phi=\pi$, and
the additional phase is $\theta=0$. The two-photon detuning $\delta$
establishes a coupling between the two effective $\Lambda$-systems. (c)
Equivalent transition diagram for the DT system at $\phi=0$ and $\theta=0$,
where the structure combines a double-$\Lambda$ system with a two-ground-state
system. The coupling between the subsystems is mediated by ground-state
coherences. }
\end{figure}

We investigate the interaction between light and matter in an atomic
ensemble using a five-level DT configuration, as illustrated in Fig.~\ref{fig:1}(a).
The atomic system comprises three ground states $|0\rangle$, $|1\rangle$,
and $|2\rangle$ along with two excited states $|A\rangle$ and $|B\rangle$.
Two weak probe fields $\varepsilon_{A}$ and $\varepsilon_{B}$  
drive the transitions $|0\rangle\leftrightarrow|A\rangle$, and $|0\rangle\leftrightarrow|B\rangle$,
respectively. In addition, four coherent control fields with Rabi
frequencies $\Omega_{A1}$, $\Omega_{A2}$, $\Omega_{B1}$, and $\Omega_{B2}$
mediate dipole-allowed transitions $|A\rangle\leftrightarrow|1\rangle$,
$|A\rangle\leftrightarrow|2\rangle$, $|B\rangle\leftrightarrow|1\rangle$,
and $|B\rangle\leftrightarrow|2\rangle$, respectively, forming a
closed-loop coherent subsystem within the DT structure. This loop
enables quantum interference and coherent control over the internal
dynamics of the atomic ensemble, making the system highly sensitive
to the relative phases of the applied control fields. The interaction
Hamiltonian of the system, expressed in units where $\hbar=1$, is
given by 
\begin{equation}
H=(-\varepsilon_{A}|A\rangle\langle0|-\varepsilon_{B}|B\rangle\langle0|-\Omega_{A1}|A\rangle\langle1|-\Omega_{A2}|A\rangle\langle2|-\Omega_{B1}|B\rangle\langle1|-\Omega_{B2}|B\rangle\langle2|)+\text{H.c.}-\delta\left(|2\rangle\langle2|-|1\rangle\langle1|\right), \label{eq:Hamiltonian}
\end{equation}

where $\delta$ represents the two-photon detuning, assigning an energy
shift of $+\delta$ to state $|1\rangle$ and $-\delta$ to $|2\rangle$.
We treat the interaction of the fields with the atoms
using the electric dipole approximation.
The detuning of the probe field appears in the full Hamiltonian as
a term of the form $\Delta|0\rangle\langle0|$; however, for simplicity
in the subsequent analysis, we omit this term in the current presentation.

To simplify the dynamics, we introduce the bright and dark states
for each subsystem coupled (bright) or decoupled (dark) from the states
$|A\rangle$ and $|B\rangle$, respectively. For the subsystem coupled
to $|A\rangle$, the bright and dark states are defined as 
\begin{align}
|\text{\ensuremath{\mathcal{B}}}_{A}\rangle & =\frac{\Omega_{A1}^{*}|1\rangle+\Omega_{A2}^{*}|2\rangle}{\Omega_{A}},\label{eq:bA}\\
|\mathcal{D}_{A}\rangle & =\frac{\Omega_{A2}|1\rangle-\Omega_{A1}|2\rangle}{\Omega_{A}},\label{eq:dA}
\end{align}
where $\Omega_{A}=\sqrt{|\Omega_{A1}|^{2}+|\Omega_{A2}|^{2}}$. Similarly,
for the subsystem coupled to $\ensuremath{|B\rangle}$, the bright
and dark states are 
\begin{align}
|\text{\ensuremath{\mathcal{B}}}_{B}\rangle & =\frac{\Omega_{B1}^{*}|1\rangle+\Omega_{B2}^{*}|2\rangle}{\Omega_{B}},\label{eq:bB}\\
|\text{\ensuremath{\mathcal{\mathcal{D}}}}_{B}\rangle & =\frac{\Omega_{B2}|1\rangle-\Omega_{B1}|2\rangle}{\Omega_{B}},\label{eq:dB}
\end{align}
with $\Omega_{B}=\sqrt{|\Omega_{B1}|^{2}+|\Omega_{B2}|^{2}}$. Here
$\Omega_{A}$ and $\Omega_{B}$ represent the effective Rabi frequencies
coupling the states $|A\rangle$ and $|B\rangle$ to their respective
bright states. By expressing the system in the new basis $\{\ensuremath{|\text{\text{\ensuremath{\mathcal{B}}}}_{A}\rangle},\ensuremath{|\text{\ensuremath{\mathcal{\mathcal{D}}}}_{A}\rangle},|\text{\text{\ensuremath{\mathcal{B}}}}_{B}\rangle,|\text{\ensuremath{\mathcal{\mathcal{D}}}}_{B}\rangle\}$,
the Hamiltonian from Eq.~\ref{eq:Hamiltonian} takes a simplified form once the dark states are decoupled
\begin{align}
H & =(-\varepsilon_{A0}|A\rangle\langle0|-\varepsilon_{B0}|B\rangle\langle0|-\Omega_{A}|A\rangle\langle\text{\ensuremath{\mathcal{B}}}_{A}|-\Omega_{B}|B\rangle\langle\text{\ensuremath{\mathcal{B}}}_{B}|)+\text{H.c.}\nonumber \\
- & \delta\left\{ C_{B}|\text{\ensuremath{\mathcal{B}}}_{A}\rangle\langle\mathcal{B}_{A}|+C_{D}|\text{\ensuremath{\mathcal{D}}}_{A}\rangle\langle\text{\ensuremath{\mathcal{D}}}_{A}|+C_{X}|\text{\text{\ensuremath{\mathcal{B}}}}_{A}\rangle\langle\text{\ensuremath{\mathcal{D}}}_{A}|+C_{X}^{*}|\text{\ensuremath{\mathcal{D}}}_{A}\rangle\langle\text{\text{\ensuremath{\mathcal{B}}}}_{A}|\right\} ,\label{eq:Hamiltonian2}
\end{align}
where the coefficients are introduced as
\begin{align}
C_{B} & =|\Omega_{A2}|^{2}-|\Omega_{A1}|^{2},\label{eq:cB} \\
C_{D} & =-C_{B},\label{eq:cD} \\
C_{X}& =-\frac{\Omega_{A1}\Omega_{A2}+\Omega_{A1}^{*}\Omega_{A2}^{*}}{|\Omega_{A1}|^{2}+|\Omega_{A2}|^{2}}.\label{eq:cX}
\end{align}

\subsection{Induced torque on DT atoms}

Let us now consider that the two weak probe fields $\varepsilon_{A}$
and $\varepsilon_{B}$ are spatially inhomogeneous, counter-propagating
beams along $z$ coordinate. These probe fields are expressed as
\begin{align}
\varepsilon_{A} & =\Omega_{A0}e^{i\Phi_{A0}(\mathbf{R})},\label{eq:fieldA0}\\
\varepsilon_{B} & =\Omega_{B0}e^{il\Phi_{B0}(\mathbf{R})},\label{eq:fieldB0}
\end{align}
where the spatial phase profiles are given by 
\begin{align}
\Phi_{A0}(\mathbf{R}) & =-l\varphi+kz,\label{eq:phaseA}\\
\Phi_{B0}(\mathbf{R}) & =-l\varphi-kz.\label{eq:phaseB}
\end{align}
 Here, $k$ is the wave number, $l$ denotes OAM or topological charge, and $\varphi$ is the azimuthal angle
in cylindrical coordinates. We consider the effects arising near the beam waist, which is situated in the plane $z = 0$.
 The amplitudes of the probe fields, corresponding
to the Rabi frequencies $\Omega_{A0}$ and $\Omega_{B0}$, are modeled
as doughnut-shaped Laguerre-Gaussian (LG) beams with a shared radial
profile  
\begin{align}
\Omega_{A0} & =|\Omega_{A0}|G(r),\label{eq:oA0}\\
\Omega_{B0} & =|\Omega_{B0}|G(r),\label{eq:oB0}
\end{align}
 where the spatial mode function is defined as $G(r)=(\frac{r}{w})^{|l|}e^{-r^{2}/w^{2}}$,
with $r$ being the radial coordinate, $w$ the beam waist, and $|\Omega_{A0}|$,$|\Omega_{B0}|$ represent the strength of the position
dependent beams.

We will derive the expression for the optical force using the semiclassical approximation.
Since only spatially varying terms in the Hamiltonian can contribute
to the force, the force arises solely from the spatial dependence
embedded in $\varepsilon_{A}$ and $\varepsilon_{B}$. Therefore,
the relevant part of the Hamiltonian for force derivation reduces
to:
\begin{equation}
H_{\mathrm{probe}}=-\varepsilon_{A}|A\rangle\langle0|-\varepsilon_{B}|B\rangle\langle0|-\varepsilon_{A}^{*}|0\rangle\langle A|-\varepsilon_{B}^{*}|0\rangle\langle B|.\label{eq:H-probe}
\end{equation}
Correspondingly, the expectation value of this part is
\begin{equation}
\langle H_{\mathrm{probe}}\rangle=-\varepsilon_{A}\sigma_{0A}-\varepsilon_{B}\sigma_{0B}-\varepsilon_{A}^{*}\sigma_{A0}-\varepsilon_{B}^{*}\sigma_{B0},\label{eq:expec. value}
\end{equation}
where $\sigma_{ij}=\langle i|\sigma|j\rangle$ represents the density matrix element in the laboratory frame.
Following the principles underlying Ehrenfest's theorem
\citep{Cook79}, the optical force is
\begin{equation}
\mathbf{F}=\sigma_{0A}\nabla\varepsilon_{A0}
+\sigma_{0B}\nabla\varepsilon_{B0}
+\sigma_{A0}\nabla\varepsilon_{A0}^{*}
+\sigma_{B0}\nabla\varepsilon_{B0}^{*}.\label{eq:ehrenfest}
\end{equation}
By substituting Eqs.~(\ref{eq:fieldA0}) and (\ref{eq:fieldB0}) into Eq.~(\ref{eq:ehrenfest})
and applying the Slowly Varying Envelope Approximation (SVEA) together
with the Rotating Wave Approximation (RWA) \citep{Scully1997}, the resulting
expression for the optical force becomes:
\begin{align}
\mathbf{F} & =i\Omega_{A0}\left(\rho_{A}^{*}-\rho_{A}\right)\nabla\Phi_{A0}+i\Omega_{B0}\left(\rho_{B}^{*}-\rho_{B}\right)\nabla\Phi_{B0}\nonumber \\
 & =2[\Omega_{A0}\text{Im}(\rho_{A})\nabla\Phi_{A0}(\mathbf{R})+\Omega_{B0}\text{Im}(\rho_{B})\nabla\Phi_{B0}(\mathbf{R})].\label{eq:force}
\end{align}
Here we ignored the dipole force related to the gradients of the field
intensities and consider only the resonance-radiation pressure related
to the gradients of the phases. The quantities $\rho_{A}$ and $\rho_{B}$
are the coherences (off-diagonal elements of the density matrix) corresponding
to the probe transitions $|0\rangle\leftrightarrow|A\rangle$, and
$|0\rangle\leftrightarrow|B\rangle$, respectively. In deriving Eq.~(\ref{eq:force})
we assume that the many-body wave function of the trapped atomic ensemble
can be approximated as a product of identical single-particle wave
functions \citep{Andersen2006}, implying that each atom experiences
the same light-induced force. Moreover, each single-atom wave function
is taken to be separable into internal and center-of-mass components.
Consequently, as shown in Eq.~(\ref{eq:force}), the force acting on
the center-of-mass coordinate $\mathbf{R}$ depends solely on the
internal-state density matrix elements.

The force expression featured in Eq.~(\ref{eq:force}) depends on the
gradients of spatial phases $\Phi_{A0}(\mathbf{R})$ and $\Phi_{B0}(\mathbf{R})$.
Applying the gradient operator in cylindrical coordinates 
\begin{equation}
\nabla\Phi=\frac{\partial\Phi}{\partial r}\hat{r}+\frac{1}{r}\frac{\partial\Phi}{\partial\varphi}\hat{\varphi}+\frac{\partial\Phi}{\partial z}\hat{z},\label{eq:operator}
\end{equation}
to $\Phi_{A0}(\mathbf{R})$ and $\Phi_{B0}(\mathbf{R})$ given in
Eqs.~(\ref{eq:phaseA}) and (\ref{eq:phaseB}) and noting that $\Phi_{A0}(\mathbf{R})$
and $\Phi_{B0}(\mathbf{R})$ have no explicit dependence on $r$,
one gets
\begin{align}
\nabla\Phi_{A0} & =\left(-\frac{l}{r}\right)\hat{\varphi}+k\hat{z},\label{eq:gr1}\\
\nabla\Phi_{B0} & =\left(-\frac{l}{r}\right)\hat{\varphi}-k\hat{z}.\label{eq:gr2}
\end{align}
Substituting these gradients into Eq.~(\ref{eq:force}), expanding
and rearranging the terms, we obtain 
\begin{equation}
\mathbf{F}=2\left[k\left(\Omega_{A0}\text{Im}(\rho_{A})-\Omega_{B0}\text{Im}(\rho_{B})\right)\hat{z}-\frac{l}{r}\left(\Omega_{A0}\text{Im}(\rho_{A})+\Omega_{B0}\text{Im}(\rho_{B})\right)\hat{\varphi}\right].\label{eq:F}
\end{equation}
To compute the torque exerted on the atomic center of mass about the
beam axis, we extract the component of the cross product of the position
vector $\mathbf{R}$ and the force vector $\mathbf{F}$ that is aligned
with the axial (beam) direction. In cylindrical coordinates, the position
vector $\ensuremath{\mathbf{R}}$ possesses only radial and axial
components, while the force $\mathbf{F}$ contains contributions in
the azimuthal and axial directions. Consequently, the only nonvanishing
contribution to the torque about the beam axis comes from the cross
product between the radial component of $\mathbf{R}$ and the azimuthal
component of $\mathbf{F}$. This yields a torque vector directed along
$\hat{z}$. Thus, the induced torque on the center of mass of the
DT atoms about the beam axis is given by
\begin{equation}
\mathbf{T}=\mathbf{r}\times\mathbf{F}=-2G^{2}(r)l\tau\,\hat{z},\label{eq:T}
\end{equation}
where we have defined the torque function 
\begin{equation}
\tau=|\Omega_{A0}|\left(\frac{|\Omega_{A0}|}{\Omega_{A0}}\text{Im}(\rho_{A})\right)+|\Omega_{B0}|\left(\frac{|\Omega_{B0}|}{\Omega_{B0}}\text{Im}(\rho_{B})\right)=\left(\frac{\text{|\ensuremath{\Omega_{A0}}|Im}(\rho_{A})+|\Omega_{B0}|\text{Im}(\rho_{B})}{G(r)}\right).\label{eq:torqueF}
\end{equation}
The expression for torque in Eq.~(\ref{eq:T}) clearly describes a
discrete, quantized torque that induces rotational motion in the atoms.
Its magnitude increases with the topological charge $l$, and the
torque exhibits a ring-like intensity profile that forms an optical
dipole trap, drawing atoms toward regions of high intensity. The torque
function given in Eq.~(\ref{eq:torqueF}) indicates its dependence
on DT system parameters through the imaginary parts of the atomic
coherences $\rho_{A}$ and $\rho_{B}$. In particular, the relative
phase among all control fields influences these coherences and thus
affects the torque on the DT atoms. For illustrative purposes, we
adopt this torque function in later sections. This phase dependence
offers a means to dynamically control the torque, enabling the generation
of a phase-sensitive, adjustable, intensity-driven circular atomic
flow. In the following section, we derive explicit analytical solutions
for the coherence terms $\rho_{A}$ and $\rho_{B}$, demonstrating
how the relative phase emerges in the equations, thereby shaping the
induced torque and influencing atomic motion within the system. 

\subsection{Optical Bloch equations}

In the DT system illustrated in Fig.~\ref{fig:1}(a), the evolution
of the probe fields and the atomic coherences is described by the
optical Bloch equations. For the optical coherences, we have \citep{Meng2014}
\begin{equation}
\frac{\partial}{\partial t}\left[\begin{array}{c}
\rho_{A}\\
\rho_{B}
\end{array}\right]=i\hat{\Omega}\left[\begin{array}{c}
\rho_{1}\\
\rho_{2}
\end{array}\right]+\left(i\Delta-\frac{\Gamma}{2}\right)\left[\begin{array}{c}
\rho_{A}\\
\rho_{B}
\end{array}\right]+i\left[\begin{array}{c}
\Omega_{A0}\\
\Omega_{B0}
\end{array}\right]\label{eq:roA,B}
\end{equation}
while for the ground-state coherences we have
\begin{equation}
\frac{\partial}{\partial t}\left[\begin{array}{c}
\rho_{1}\\
\rho_{2}
\end{array}\right]=i\hat{\Omega}^{\dagger}\left[\begin{array}{c}
\rho_{A}\\
\rho_{B}
\end{array}\right]+i\left(\Delta\hat{I}+\hat{\delta}\right)\left[\begin{array}{c}
\rho_{1}\\
\rho_{2}
\end{array}\right],\label{eq:rho1,2}
\end{equation}
where we define the field interaction matrix $\hat{\Omega}$ and the
two-photon detuning matrix $\hat{\delta}$ as
\begin{equation}
\hat{\Omega}=\left[\begin{array}{cc}
\Omega_{A1} & \Omega_{A2}\\
\Omega_{B1} & \Omega_{B2}
\end{array}\right]\,,\qquad\hat{\delta}=\left[\begin{array}{cc}
\delta & 0\\
0 & -\delta
\end{array}\right],\label{eq:matrices}
\end{equation}
and $\hat{I}$ is identity matrix. In these equations, $\rho_{A}$
and $\rho_{B}$ represent the optical coherences corresponding to
the transitions $|0\rangle\leftrightarrow|A\rangle$ and $|0\rangle\leftrightarrow|B\rangle$,
respectively (which appear in Eq.~(\ref{eq:torqueF}) for the induced
torque), whereas $\rho_{1}$ and $\rho_{2}$ denote the ground-state
coherences between $|0\rangle$ and $|1\rangle$ as well as $|0\rangle$
and $|2\rangle$. The parameter $\Gamma$ represents the spontaneous
decay rate of the excited states, while $\delta$ and $\Delta$ correspond
to the two-photon detuning and probe detuning, respectively. 

In what follows, we focus on the case where the four control fields
have equal amplitudes $\Omega$, allowing the complex Rabi frequencies
to be written as
\begin{equation}
\Omega_{i}=\Omega e^{i\phi_{i}},\qquad i=A1,A2,B1,B2,\label{eq:Omegas}
\end{equation}
with $\phi_{i}$ representing the phase of each field. To derive the optical Bloch
equations, we assume that the probe fields are significantly weaker
than the control fields, ensuring that the atomic population remains
predominantly in the ground state $|0\rangle$. This allows us to
treat the probe fields as a perturbation. Furthermore, all rapidly
oscillating exponential factors related to the central frequencies
and wave vectors have been removed, leaving only the slowly varying
amplitudes. This is justified under the SWEA and RWA, which assume that the field envelopes vary slowly in space and time compared to the optical wavelength and period, so that the rapidly oscillating terms average to zero.

We seek the steady-state solutions of the probe fields $\Omega_{A0},\Omega_{B0}$
by dropping the time derivatives in the equations (\ref{eq:roA,B})
and (\ref{eq:rho1,2}) (i.e., when $\frac{\partial}{\partial t}\rightarrow0$).
This corresponds to the long-term behavior of the optical coherences
$\rho_{A}$ and $\rho_{B}$. From Eq.~(\ref{eq:rho1,2}), we obtain
\begin{equation}
\left[\begin{array}{c}
\rho_{1}\\
\rho_{2}
\end{array}\right]=-\left(\Delta\hat{I}+\hat{\delta}\right)^{-1}\hat{\Omega}^{\dagger}\left[\begin{array}{c}
\rho_{A}\\
\rho_{B}
\end{array}\right].\label{eq:ro1,2S}
\end{equation}
Substituting this expression into Eq.~(\ref{eq:roA,B}) yields
\begin{equation}
\hat{\Omega}\left(\Delta\hat{I}+\hat{\delta}\right)^{-1}\hat{\Omega}^{\dagger}\left[\begin{array}{c}
\rho_{A}\\
\rho_{B}
\end{array}\right]-\left(\Delta+i\frac{\Gamma}{2}\right)\left[\begin{array}{c}
\rho_{A}\\
\rho_{B}
\end{array}\right]=\left[\begin{array}{c}
\Omega_{A0}\\
\Omega_{B0}
\end{array}\right].\label{eq:sols}
\end{equation}
Rearranging Eq.~(\ref{eq:sols}) we obtain the steady-state solutions
for $\rho_{A}$ and $\rho_{B}$ 
\begin{equation}
\left[\begin{array}{c}
\rho_{A}\\
\rho_{B}
\end{array}\right]=\left[\hat{\Omega}\left(\Delta\hat{I}+\hat{\delta}\right)^{-1}\hat{\Omega}^{\dagger}-\left(\Delta+i\frac{\Gamma}{2}\right)\hat{I}\right]^{-1}\left[\begin{array}{c}
\Omega_{A0}\\
\Omega_{B0}
\end{array}\right].\label{eq:solutions}
\end{equation}


\section{Phase sensitive torque}

Eqs.~(\ref{eq:torqueF}), (\ref{eq:matrices}), and (\ref{eq:solutions}) collectively indicate that the variation
of torque function $\tau$ depends on the Rabi frequency $\Omega$
as well as the two-photon detuning $\delta$. However, these expressions
do not explicitly reveal the dependence on the relative phase $\phi$ between
the applied fields. In the following, we will show this phase dependence
by considering different cases of light-matter interaction.

\subsection{Nonzero Two-Photon Detuning ($\delta\protect\neq0$)}

For nonzero two photon detuning $\delta\neq0$ and equal control field amplitudes $\Omega$,
we simplify the inverse term in Eq.~(\ref{eq:solutions}) using Eqs.~(\ref{eq:matrices})
and (\ref{eq:Omegas}) as follows
\begin{equation}
\hat{\Omega}\left(\Delta\hat{I}+\hat{\delta}\right)^{-1}\hat{\Omega}^{\dagger}=\frac{2\Omega^{2}\Delta}{(\Delta^{2}-\delta^{2})}\left[\begin{array}{cc}
1 & e^{i\left(\theta+\frac{\phi}{2}\right)}\left(\cos\left(\frac{\phi}{2}\right)-i\frac{\delta}{\Delta}\sin\left(\frac{\phi}{2}\right)\right)\\
e^{-i\left(\theta+\frac{\phi}{2}\right)}\left(\cos\left(\frac{\phi}{2}\right)+i\frac{\delta}{\Delta}\sin\left(\frac{\phi}{2}\right)\right) & 1
\end{array}\right],\label{eq:15}
\end{equation}
where $\phi=(\phi_{A1}-\phi_{B1})-\theta$ and $\theta=\phi_{A2}-\phi_{B2}$.
Here, $\phi$ represents the total relative phase among all four
control fields, and $\theta$ provides an additional (phase) degree
of control over the optical torque exerted on the DT system. Substititng
Eq.~(\ref{eq:15}) into Eq.~(\ref{eq:solutions}) yields the optical
coherence vector 
\begin{align}
\left[\begin{array}{c}
\rho_{A}\\
\rho_{B}
\end{array}\right] & =\frac{\frac{(\Delta^{2}-\delta^{2})}{2\Omega^{2}\Delta^{2}}}{\left(1-\frac{(\Delta^{2}-\delta^{2})(2\Delta+i\Gamma)}{4\Omega^{2}\Delta}\right)^{2}-\cos^{2}\left(\frac{\phi}{2}\right)-\frac{\delta^{2}}{\Delta^{2}}\sin^{2}\left(\frac{\phi}{2}\right)}\nonumber \\
 & \times\left[\begin{array}{cc}
\Delta-\frac{(\Delta^{2}-\delta^{2})(2\Delta+i\Gamma)}{4\Omega^{2}} & -\Delta e^{i\left(\theta+\frac{\phi}{2}\right)}\left(\cos\left(\frac{\phi}{2}\right)-i\frac{\delta}{\Delta}\sin\left(\frac{\phi}{2}\right)\right)\\
-\Delta e^{-i\left(\theta+\frac{\phi}{2}\right)}\left(\cos\left(\frac{\phi}{2}\right)+i\frac{\delta}{\Delta}\sin\left(\frac{\phi}{2}\right)\right) & \Delta-\frac{(\Delta^{2}-\delta^{2})(2\Delta+i\Gamma)}{4\Omega^{2}}
\end{array}\right]\left[\begin{array}{c}
\Omega_{A0}\\
\Omega_{B0}
\end{array}\right].\label{eq:sol16}
\end{align}

\subsubsection{Relative phase $\phi=\pi$}

A particularly instructive scenario arises when the applied fields
have a relative phase of $\pi$. By setting the phases of individual
control fields to $(\phi_{A1},\phi_{B1},\phi_{A2},\phi_{B2})=(\pi,0,0,0)$
we immediately have $\phi=\pi$ and $\theta=0$. In this case, the
system undergoes a reorganization due to phase-controlled interference.
The $\pi$-phase difference rotates the bright-dark basis
in Eqs.~(\ref{eq:bA})--(\ref{eq:dB}) into symmetric and antisymmetric
superpositions of
\begin{align}
|\text{\ensuremath{\mathcal{B}}}_{A}\rangle & =\frac{\Omega e^{i\pi}\lvert1\rangle+\Omega\lvert2\rangle}{\sqrt{2}\Omega}=\frac{|2\rangle-|1\rangle}{\sqrt{2}}\rightarrow|-\rangle,\label{eq:da}\\
|\text{\text{\ensuremath{\mathcal{D}}}}_{A}\rangle & =\frac{|1\rangle+|2\rangle}{\sqrt{2}}\rightarrow|+\rangle\label{eq:bb}\\
|\text{\text{\ensuremath{\mathcal{B}}}}_{B}\rangle & =\frac{\Omega|1\rangle+\Omega|2\rangle}{\sqrt{2}\Omega}=\frac{|1\rangle+|2\rangle}{\sqrt{2}}\rightarrow|+\rangle\label{eq:dd}\\
|\text{\text{\ensuremath{\mathcal{D}}}}_{B}\rangle & =\frac{|1\rangle-|2\rangle}{\sqrt{2}}\rightarrow|-\rangle.\label{eq:db}
\end{align}
Note that the \textquotedblleft flip\textquotedblright{} between $\frac{|2\rangle-|1\rangle}{\sqrt{2}}$
and $\frac{|1\rangle-|2\rangle}{\sqrt{2}}$ is just an unobservable
global minus sign, thus both conventions represent the same antisymmetric
state. 

In addition, Eqs.~(\ref{eq:cB})--(\ref{eq:cX}) yield $C_{B}=C_{D}=0$
and $C_{X}=1$, simplifying the Hamiltonian in Eq.(\ref{eq:Hamiltonian2})
to
\begin{equation}
H=(-\varepsilon_{A}|A\rangle\langle0|-\varepsilon_{B}|B\rangle\langle0|-\Omega_{A}|A\rangle\langle-|-\Omega_{B}|B\rangle\langle+|)+\text{H.c.}-\delta\left(|+\rangle\langle-|+|-\rangle\langle+|\right).\label{eq:Hamiltonian3}
\end{equation}
This indicates that the applied $\pi$-phase difference leads to coupling
between $|A\rangle$ and the antisymmetric $|-\rangle$ state, and between
$|B\rangle$ and the symmetric state $|+\rangle$. The control fields
driving these transitions are renormalized into effective couplings
with Rabi frequencies $\Omega_{A}=\Omega_{B}=\sqrt{2}\Omega$, a result
of constructive interference in the transformed basis. This transformation
reconfigures the DT system into two coupled $\Lambda$ subsystems
(Fig.~\ref{fig:1}~(b)). The first $\Lambda$ subsystem involves
transitions $|0\rangle\leftrightarrow|A\rangle\leftrightarrow|-\rangle$
, while the second involves $|0\rangle\leftrightarrow|B\rangle\leftrightarrow|+\rangle$.
The term $\delta$ introduces cross-talk between two $\Lambda$ subsystems. 

\begin{figure}
\includegraphics[width=0.7\columnwidth]{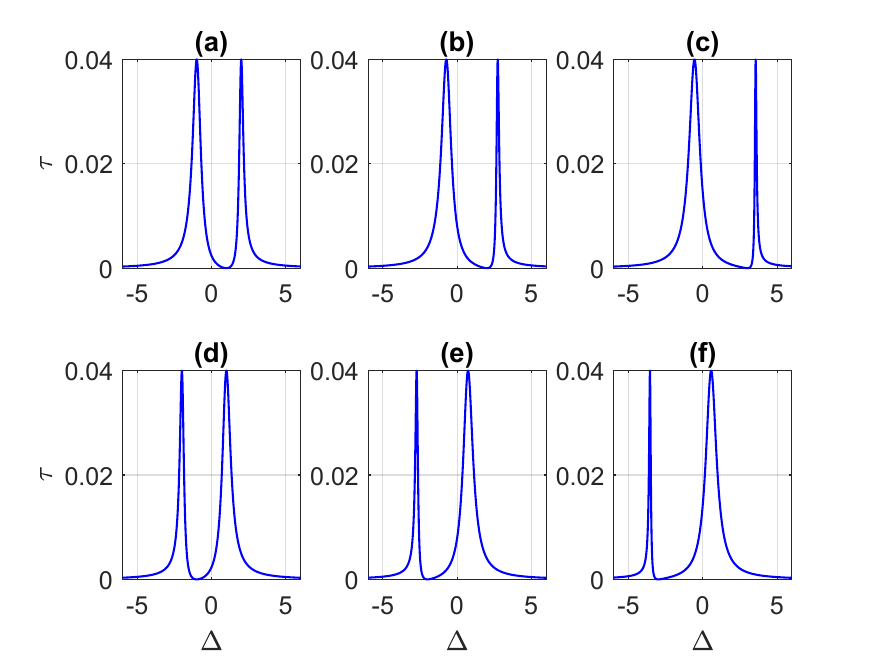}

\caption{\label{fig:2} The induced torque function $\tau$ as a function of probe
detuning $\Delta$ at nonzero two-photon detuning. Subplots (a), (b),
and (c) correspond to $\delta=\Gamma,2\Gamma,$ and $3\Gamma$, while
subplots (d), (e), and (f) correspond to $\delta=-\Gamma,-2\Gamma,$ and
$-3\Gamma$. The strength of probe fields are $|\Omega_{A0}|=|\Omega_{B0}|=0.1\Gamma$.
The amplitudes of the individual control fields are $|\Omega_{A1}|=|\Omega_{A2}|=|\Omega_{B1}|=|\Omega_{B2}|=|\Omega|=\Gamma$.
The phases of the individual control fields are set as $(\phi_{A1},\phi_{B1},\phi_{A2},\phi_{B2})=(\pi,0,0,0)$
leading to $\phi=\pi$ and $\theta=0$, which simplifies the DT system
to two coupled $\Lambda$ systems. All the relevant parameters are
scaled with $\Gamma$. }
\end{figure}

Under this condition Eq.~(\ref{eq:sol16}) simplifies to 
\begin{equation}
\left[\begin{array}{c}
\rho_{A}\\
\rho_{B}
\end{array}\right]=\frac{8\Omega^{2}}{\left[4\Omega^{2}-(2\Delta+i\Gamma)(\Delta+\delta)\right]\left[4\Omega^{2}-(2\Delta+i\Gamma)(\Delta-\delta)\right]}\left[\begin{array}{cc}
\Delta-\frac{(\Delta^{2}-\delta^{2})(2\Delta+i\Gamma)}{4\Omega^{2}} & -\delta\\
-\delta & \Delta-\frac{(\Delta^{2}-\delta^{2})(2\Delta+i\Gamma)}{4\Omega^{2}}
\end{array}\right]\left[\begin{array}{c}
\Omega_{A0}\\
\Omega_{B0}
\end{array}\right].\label{eq:sol17}
\end{equation}

The off-diagonal term $-\delta$ explicitly couples the two probe
fields $\Omega_{A0}$ and $\Omega_{B0}$. This two-photon detuning-mediated
interaction leads to a phase-sensitive coupling, which directly affects
both the optical response of the system and its mechanical dynamics. 

Figure~\ref{fig:2} illustrates the induced torque $\tau$ as a
function of the probe detuning $\Delta$ for different values of $\delta$:
$\delta=\Gamma,2\Gamma,3\Gamma$ in panels (a), (b), and (c), and
$\delta=-\Gamma,-2\Gamma,-3\Gamma$ in panels (d), (e), and (f), respectively.
In all cases, we consider equal probe amplitudes: $|\Omega_{A0}|=|\Omega_{B0}|$.
As defined in Eq.~\eqref{eq:torqueF}, the torque function corresponds
to the sum of the imaginary parts of the coherences $\rho_{A}$ and
$\rho_{B}$, which are associated with the absorption of the probe
fields $\Omega_{A0}$ and $\Omega_{B0}$. Consequently, the torque
profile resembles the absorption spectrum of the probe fields. As
shown in Fig.~\ref{fig:2}, the torque exhibits peaks on either
side of $\Delta=\delta$, which corresponds to the condition where
the system exhibits EIT for both probe fields. According to Eq.~\eqref{eq:sol17},
when $|\Omega_{A0}|=|\Omega_{B0}|$, the coherence terms $\rho_{A}$
and $\rho_{B}$ (and thus their imaginary parts) vanish at $\Delta=\delta$,
leading to transparency. Adjusting the two-photon detuning $\delta$
shifts the EIT window for both probes and thereby changes the position
and width of the torque peaks. For instance, when $\delta=\Gamma$,
EIT occurs at $\Delta=\Gamma$ (Fig.~\ref{fig:2}(a)), while for
$\delta=-\Gamma$, the transparency window shifts to $\Delta=-\Gamma$
(Fig.~\ref{fig:2}(d)). Increasing $\delta$ to larger positive
(negative) values gradually broadens the left (right) torque peak.
This behavior confirms that the torque is rooted in absorptive processes:
when EIT suppresses absorption, the torque vanishes; when absorption
is enhanced near resonance, the torque reaches its peak and reflects
the transfer of OAM from the twisted light
fields to the DT atoms. 

\begin{figure}
\includegraphics[width=0.7\columnwidth]{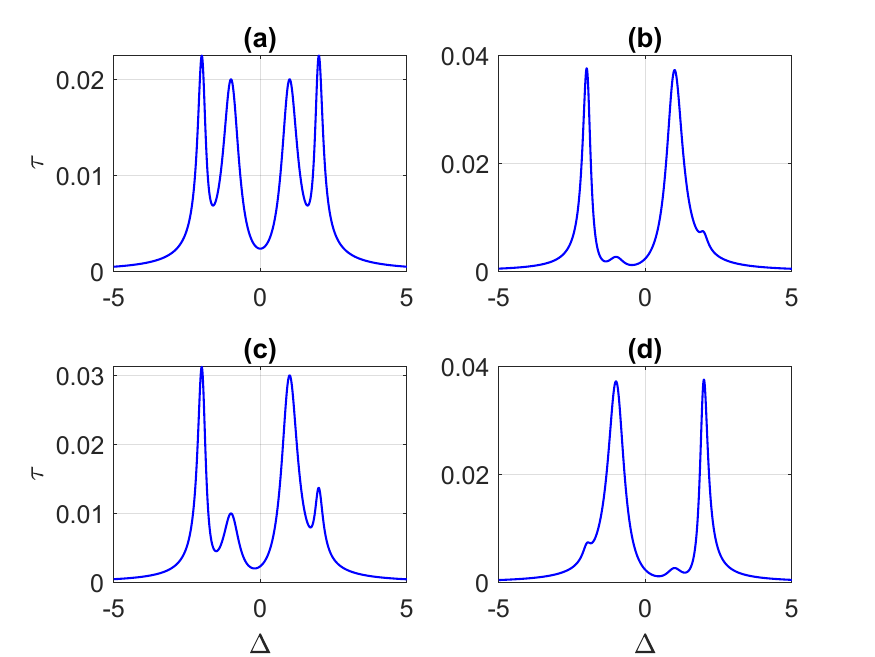}

\caption{\label{fig:3} The induced  torque function $\tau$ as a function of probe
detuning $\Delta$ at nonzero two-photon detuning ($\delta=\Gamma$)
and for different values of additional phase $\theta$. The strength
of probe fields are $|\Omega_{A0}|=|\Omega_{B0}|=0.1\Gamma$. The
amplitudes of the individual control fields are $|\Omega_{A1}|=|\Omega_{A2}|=|\Omega_{B1}|=|\Omega_{B2}|=|\Omega|=\Gamma$.
The phases of the individual control fields are set as $(\phi_{A1},\phi_{B1},\phi_{A2},\phi_{B2})=(\frac{\pi}{2},0,0,\frac{\pi}{2})$
leading to $\phi=\pi$ and $\theta=-\pi/2$ (a), $(\phi_{A1},\phi_{B1},\phi_{A2},\phi_{B2})=(\frac{\pi}{6},0,0,\frac{5\pi}{6})$
leading to $\phi=\pi$ and $\theta=-5\pi/6$ (b), $(\phi_{A1},\phi_{B1},\phi_{A2},\phi_{B2})=(\frac{\pi}{3},0,0,\frac{2\pi}{3})$
leading to $\phi=\pi$ and $\theta=-2\pi/3$ (c), and $(\phi_{A1},\phi_{B1},\phi_{A2},\phi_{B2})=(\frac{5\pi}{6},0,0,\frac{\pi}{6})$
leading to $\phi=\pi$ and $\theta=-\pi/6$ (d).}
\end{figure}

To isolate the role of the emergent phase $\theta$ on the induced
torque, we fix $\phi=\pi$ and vary $\theta$ by adjusting the control
field phases. Specifically, we consider four different phase configurations
for the control fields: 
\begin{equation}
\begin{array}{ll}
\text{(a):} & (\phi_{A1},\phi_{B1},\phi_{A2},\phi_{B2})=(\frac{\pi}{2},0,0,\frac{\pi}{2})\Rightarrow\phi=\pi,\quad\theta=-\pi/2,\\[1mm]
\text{(b):} & (\phi_{A1},\phi_{B1},\phi_{A2},\phi_{B2})=(\frac{\pi}{6},0,0,\frac{5\pi}{6})\Rightarrow\phi=\pi,\quad\theta=-5\pi/6,\\[1mm]
\text{(c):} & (\phi_{A1},\phi_{B1},\phi_{A2},\phi_{B2})=(\frac{\pi}{3},0,0,\frac{2\pi}{3})\Rightarrow\phi=\pi,\quad\theta=-2\pi/3,\\[1mm]
\text{(d):} & (\phi_{A1},\phi_{B1},\phi_{A2},\phi_{B2})=(\frac{5\pi}{6},0,0,\frac{\pi}{6})\Rightarrow\phi=\pi,\quad\theta=-\pi/6.
\end{array}\label{eq:phis}
\end{equation}
Despite $\phi$ being constant, variations in $\theta$ redistribute
the torque variation (Fig.~\ref{fig:3}), revealing $\theta$ as
an independent control parameter. From Eq.~(\ref{eq:sol16}), the off-diagonal
elements of the response matrix governing $\rho_{A}$ and $\rho_{B}$
include terms of the form $e^{\pm i(\theta+\frac{\phi}{2})}$. Therefore,
even when with $\phi=\pi$, tuning $\theta$ adjusts the phase difference
between the $|A\rangle$- and $|B\rangle$-mediated pathways, thereby
altering the interference conditions. This phase-dependent interference
directly impacts the imaginary parts of the coherences $\rho_{A}$
and $\rho_{B}$, which are responsible for the torque generation.
For instance, the case $\theta=-\frac{\pi}{2}$ (Fig.~\ref{fig:3}(a))
results in four pronounced torque peaks, reflecting constructive interference
between the two pathways. 

\subsubsection{Relative phase $\phi=0$}

Next, we consider the case where all control field phases are set
to zero $(\phi_{A1},\phi_{B1},\phi_{A2},\phi_{B2})=(0,0,0,0)$, yielding
$\phi=0,\,\theta=0$. From Eqs.~(\ref{eq:bA})--(\ref{eq:dB}) it follows
that the bright states $\lvert\text{\text{\ensuremath{\mathcal{B}}}}_{A}\rangle$
and $\lvert\text{\text{\ensuremath{\mathcal{B}}}}_{B}\rangle$ collapse
into a single symmetric superposition 
\begin{equation}
\lvert\text{\text{\ensuremath{\mathcal{B}}}}_{A}\rangle=\lvert\text{\text{\ensuremath{\mathcal{B}}}}_{B}\rangle=\frac{\lvert1\rangle+\lvert2\rangle}{\sqrt{2}}\rightarrow\lvert+\rangle,\label{eq:+}
\end{equation}
while the dark states become $|\text{\text{\ensuremath{\mathcal{D}}}}_{A}\rangle,|\text{\text{\ensuremath{\mathcal{D}}}}_{B}\rangle\rightarrow\lvert-\rangle$.
In this transformed basis, the Hamiltonian becomes
\begin{equation}
H=(-\varepsilon_{A}\,|A\rangle\langle0|-\varepsilon_{B}\,|B\rangle\langle0|-\Omega_{A}\,|A\rangle\langle+|-\Omega_{B}\,|B\rangle\langle+|)+\text{H.c.}+\delta\left\{ |+\rangle\langle-|+|-\rangle\langle+|\right\} .\label{eq:Hamiltonian4}
\end{equation}
 This configuration couples both excited states $\lvert A\rangle$
and $\lvert B\rangle$ to the same symmetric ground-state superposition
$\lvert+\rangle$, forming two $\Lambda$ subsystems: $\,|0\rangle\leftrightarrow\,|A\rangle\leftrightarrow\,|+\rangle$
and $|0\rangle\leftrightarrow\,|B\rangle\leftrightarrow\,|+\rangle$.
As a result, the DT configuration effectively reduces to a double-$\Lambda$
(DL) structure (Fig.~\ref{fig:1}~(c)), with coherent coupling
between the DL and the antysymetric state $\lvert-\rangle$ mediated
by the two-photon detuning $\delta$.

Equation~(\ref{eq:sol16}) now takes the form
\begin{equation}
\left[\begin{array}{c}
\rho_{A}\\
\rho_{B}
\end{array}\right]=\frac{-2}{\Delta(2\Delta+i\Gamma)\left(2-\frac{(\Delta^{2}-\delta^{2})(2\Delta+i\Gamma)}{4\Omega^{2}\Delta}\right)}\left[\begin{array}{cc}
\Delta-\frac{(\Delta^{2}-\delta^{2})(2\Delta+i\Gamma)}{4\Omega^{2}} & -\Delta\\
-\Delta & \Delta-\frac{(\Delta^{2}-\delta^{2})(2\Delta+i\Gamma)}{4\Omega^{2}}
\end{array}\right]\left[\begin{array}{c}
\Omega_{A0}\\
\Omega_{B0}
\end{array}\right].\label{DT}
\end{equation}

\begin{figure}
\includegraphics[width=0.7\columnwidth]{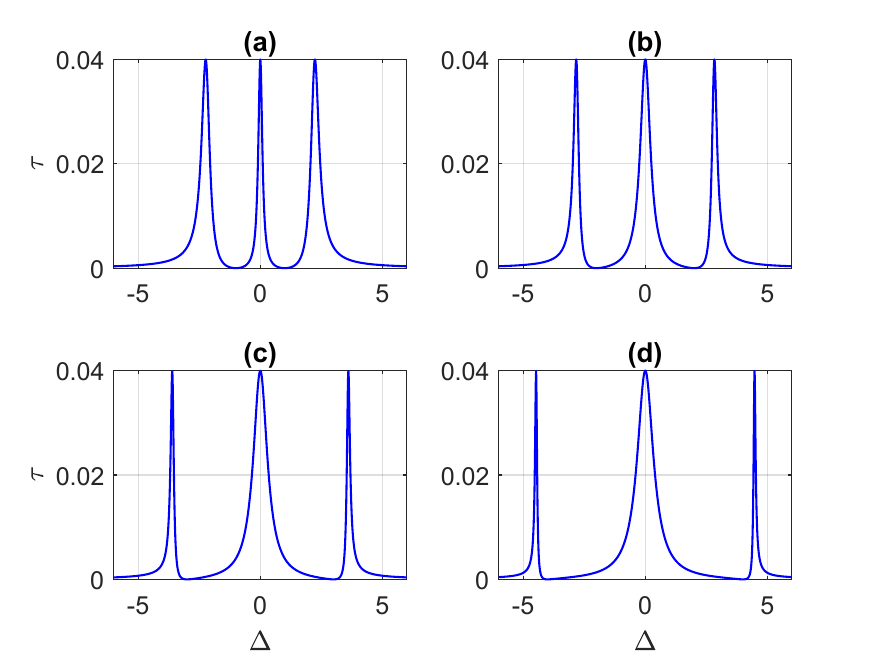}

\caption{\label{fig:4} The induced torque function $\tau$ as a function of probe
detuning $\Delta$ at nonzero two-photon detuning ($\delta=\Gamma$).
The strength of probe fields are $|\Omega_{A0}|=|\Omega_{B0}|=0.1\Gamma$.
The amplitudes of the individual control fields are $|\Omega_{A1}|=|\Omega_{A2}|=|\Omega_{B1}|=|\Omega_{B2}|=|\Omega|=\Gamma$.
Subplots (a), (b), (c) and (d) correspond to $\delta=\Gamma,2\Gamma,3\Gamma$
and $4\Gamma$. The phases of the individual control fields are set
as $(\phi_{A1},\phi_{B1},\phi_{A2},\phi_{B2})=(0,0,0,0)$ leading
to $\phi=\theta=0$, which simplifies the DT system to a DL system
combined with two ground levels with no light (degenerate DT).}
\end{figure}

The torque function $\tau$ is plotted as a function of probe detuning
$\Delta$ for different value of $\delta=\Gamma,2\Gamma,3\Gamma$ and
$4\Gamma$ in Fig.~\ref{fig:4}. The system now exhibits three distinct
peaks in the torque profile, which surround two EIT channels located
at $\Delta=\pm\delta$ for both probe fields. This behavior follows
directly from the diagonal terms of the coherence matrix in Eq.~(\ref{DT}),
which vanish when $|\Omega_{A0}|=|\Omega_{B0}|$ and the probe detuning
satisfies $\Delta^{2}-\delta^{2}=0$. Increasing the two-photon detuning
$\delta$ broadens the central torque peak while sharpening the two lateral peaks. This is because the EIT windows shift further apart as $\delta$
increases, occurring at $\pm\delta=\Gamma,2\Gamma,3\Gamma$ and $4\Gamma$.
These peaks correspond to enhanced OAM transfer from the helical phase
gradients of the LG probe fields to atoms.

\subsubsection{Intermediate relative phases $0<\phi<\pi$ }

\begin{figure}
\includegraphics[width=0.7\columnwidth]{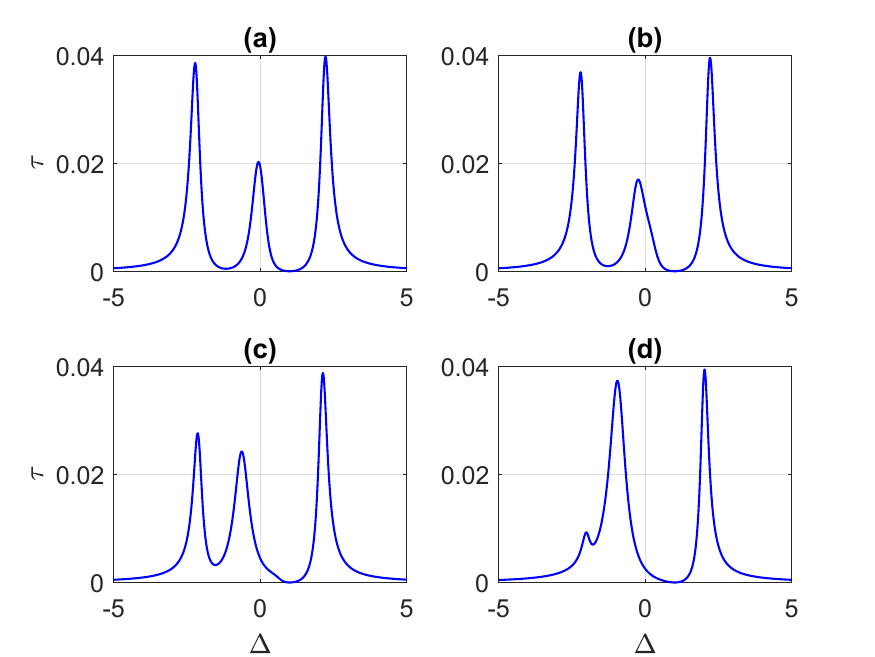}

\caption{\label{fig:5}The induced torque function $\tau$ as a function of probe detuning $\Delta$,
evaluated at nonzero two-photon detuning $\delta=\Gamma$ for different
values of the relative phase $\phi$. The strengths of the probe fields
are $|\Omega_{A0}|=|\Omega_{B0}|=0.1\Gamma$, and the amplitudes of
the individual control fields are $|\Omega_{A1}|=|\Omega_{A2}|=|\Omega_{B1}|=|\Omega_{B2}|=|\Omega|=\Gamma$.
The phases of the control fields $(\phi_{A1},\phi_{B1},\phi_{A2},\phi_{B2})$
are set as follows: (a) $(\frac{\pi}{6},0,0,0)$, (b) $(\frac{\pi}{4},0,0,0)$,
(c) $(\frac{\pi}{2},0,0,0)$, and (d) $(\frac{5\pi}{6},0,0,0)$, corresponding
to different values of $\phi=\phi_{A1}$ with $\theta=0$.}
\end{figure}

Figure~\ref{fig:5} illustrates the influence of the intermediate
relative phase values $\phi$ ($0<\phi<\pi$ ) on the torque spectrum
for a fixed $\theta=0$. Unlike the symmetric torque profile observed
for $\phi=\pi$ (Fig.~\ref{fig:2}) and $\phi=0$ (Fig.~\ref{fig:4}),
a nonzero $\phi$ introduces a pronounced asymmetry. As $\phi$ increases,
the left torque peak gradually diminishes and eventually
merges with the central peak, while the right peak remains
relatively stable. This asymmetry arises from phase-sensitive interference
in the coherence terms $\rho_{A}$ and $\rho_{B}$ given in Eq.~(\ref{eq:sol16}).
Their imaginary parts, which govern the torque response, include complex
phase-dependent factors of the form $e^{\pm i\frac{\phi}{2}}\left(\cos\frac{\phi}{2}\mp i\frac{\delta}{\Delta}\sin\frac{\phi}{2}\right)$.
These expressions show that variations in $\phi$ modulate both the
relative amplitudes and phases of the atomic coherences. As a result, the interference conditions are altered across the spectrum, leading
to a redistribution of peak intensities and the emergence of the observed
torque asymmetry.

\subsection{Zero Two-Photon Detuning ($\delta=0$)}

Thus far we have considered the case of nonzero two-photon detuning. Let
us now consider a situation when the two-photon detuning is zero $\delta=0$.
For $\phi=\pi$, the coherence matrix simplifies considerably
\begin{equation}
\left[\begin{array}{c}
\rho_{A}\\
\rho_{B}
\end{array}\right]=\frac{8\Omega^{2}}{\left[4\Omega^{2}-\Delta(2\Delta+i\Gamma)\right]^{2}}\left[\begin{array}{cc}
\Delta-\frac{\Delta^{2}(2\Delta+i\Gamma)}{4\Omega^{2}} & 0\\
0 & \Delta-\frac{\Delta^{2}(2\Delta+i\Gamma)}{4\Omega^{2}}
\end{array}\right]\left[\begin{array}{c}
\Omega_{A0}\\
\Omega_{B0}
\end{array}\right].\label{eq:coherencesd0-2}
\end{equation}

In this case, the off-diagonal terms in the coherence matrix vanish, and the
DT system reduces to a situation where the two probe fields are completely
decoupled. Under these conditions, the DT configuration behaves as
two independent $\Lambda$-type systems (or, equivalently, as two
uncoupled EIT channels). The $\pi$-phase difference in the control
fields reorganizes the ground states $|1\rangle$, and $|2\rangle$
into the orthogonal superpositions $|+\rangle=\frac{|1\rangle+|2\rangle}{\sqrt{2}}$
and $|-\rangle=\frac{|1\rangle-|2\rangle}{\sqrt{2}}$. Thus, each
probe field interacts with its respective superposition state, splitting
the system into two distinct $\Lambda$ subsystems. 

For $\delta=0$, setting $\phi_{A1}=\phi_{B1}=\phi_{A2}=\phi_{B2}=0$
, results in the overall relative phase $\phi=\theta=0$. In this
case, the coherence terms take the form
\begin{equation}
\left[\begin{array}{c}
\rho_{A}\\
\rho_{B}
\end{array}\right]=\frac{-2}{\Delta(2\Delta+i\Gamma)\left(2-\frac{\Delta(2\Delta+i\Gamma)}{4\Omega^{2}}\right)}\left[\begin{array}{cc}
\Delta-\frac{\Delta^{2}(2\Delta+i\Gamma)}{4\Omega^{2}} & -\Delta\\
-\Delta & \Delta-\frac{\Delta^{2}(2\Delta+i\Gamma)}{4\Omega^{2}}
\end{array}\right]\left[\begin{array}{c}
\Omega_{A0}\\
\Omega_{B0}
\end{array}\right].\label{eq:coherencesd0-1}
\end{equation}

\begin{figure}
\includegraphics[width=0.7\columnwidth]{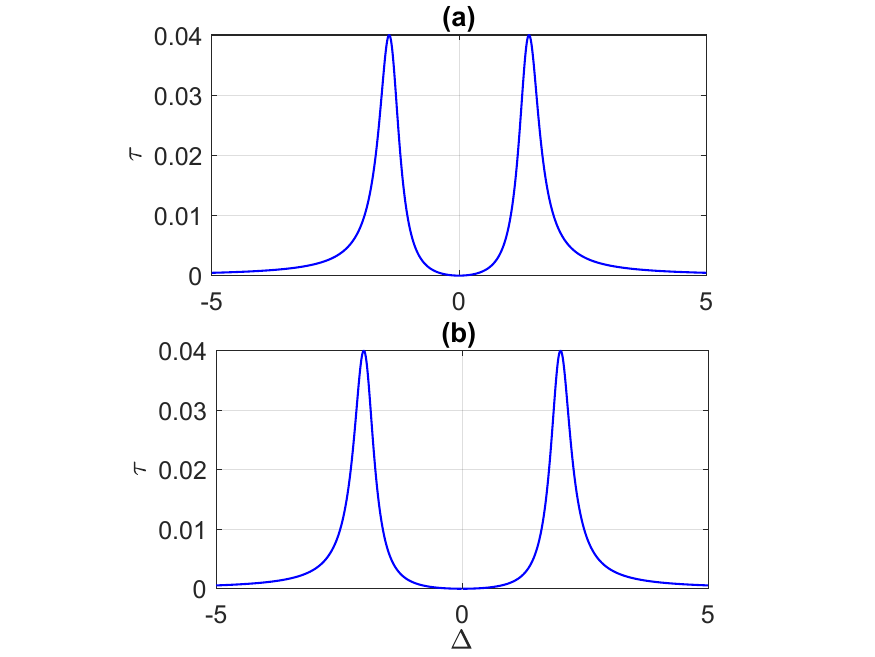}

\caption{\label{fig:6} The torque function $\tau$ as a function of probe
detuning $\Delta$ at zero two-photon detuning ($\delta=0$) and for
different values of relative phase $\phi$. The strength of probe
fields are $|\Omega_{A0}|=|\Omega_{B0}|=0.1\Gamma$. The amplitudes
of the individual control fields are $|\Omega_{A1}|=|\Omega_{A2}|=|\Omega_{B1}|=|\Omega_{B2}|=|\Omega|=\Gamma$.
The phases of the individual control fields are set as $\phi_{B1}=\phi_{A2}=\phi_{B2}=0,$
(a) $\phi_{A1}=\pi$ and (b) $0$ (leading to $\phi=\phi_{A1}$ and
$\theta=0$).}
\end{figure}

This corresponds to a symmetric (degenerate) double-$\Lambda$ configuration.
The nonzero off-diagonal terms in the coherence matrix introduce coupling
between the two probe fields. This coupling induces interference between
the imaginary parts of $\rho_{A}$ and $\rho_{B}$ which govern the
torque. The resulting torque variations are illustrated in Fig.$~\ref{fig:6}$. The torque exhibits an absorption-like profile: it vanishes
at the center due to EIT, while reaching maxima on either side as
a result of enhanced absorption. In the case of $\phi=\pi$, shown
in Fig.$~\ref{fig:6}$(a), the absence of probe coupling suppresses
the joint interaction between the two probe fields. This modifies
the spectral torque response, leading to a reduced separation between
the torque extrema. 

Before concluding, we note that in optical beams carrying OAM, a macroscopic particle can experience a spinning torque that causes it to rotate about its own axis \citep{Yanhui2023NC,Xiaohao2024NC}. In contrast, in our model each atom is treated as a point-like particle on which the force acts individually. Therefore, a spinning torque on a single atom is not defined. Our analysis considers only the orbital torque, which induces rotational motion of the atomic center of mass around the beam axis. Since the atomic ensemble is not treated as a rigid body, no collective spinning motion is expected.

\section{Summary}

We theoretically demonstrate phase-controllable optical torque in
a five-level double tripod (DT) atomic system driven by four strong coherent
control fields and two weak optical vortex probe fields carrying orbital
angular momentum (OAM). The spatial phase gradients of the OAM fields
induce a quantized torque, transferred to atoms to drive rotational
motion and directed atomic flow in an annular geometry. By analytically
solving the optical Bloch equations in steady state, we reveal that
both the torque and resultant rotation exhibit high phase sensitivity,
arising from coherent reconfigurations of the system into either coupled-$\Lambda$
or double-$\Lambda$ configurations depending on the relative phases
($\phi,\theta$) of the control fields. Each configuration displays
distinct torque characteristics. The multilevel DT configuration explored here represents a highly flexible setup, which allows employing several degrees of freedom to get different torque effects not available using simpler systems. Indeed, here the control over the total relative phase ($\phi$), extra phase ($\theta$), and two-photon detuning ($\delta$) can be used, affecting the overall torque behavior and rotation dynamics. Moreover, we emphasize that the observed toroidal atomic currents and related annular geometry effects are specifically related to the annular intensity distribution of the vortex probe beams. This phase-controlled light-induced torque establishes a versatile mechanism for optical manipulation
of quantized mechanical action in multilevel atomic ensembles. 

In the present model, the control fields are assumed to be spatially uniform, and the optical torque arises from the spatial phase gradients of the probe fields carrying OAM. Recent studies have demonstrated that spatial variations in the field amplitude can significantly influence Rabi oscillations and spin–orbit coupling effects in structured light fields \citep{Guohua2023,Zhang2024}. This indicates that the induced torque could, in principle, be controlled through both the spatial amplitude and phase profiles of a vortex control field. If the control fields were also spatially structured in our DT system, their amplitude distribution and OAM phase could introduce additional torque components. However, not all control fields can simultaneously carry vortices in the DT configuration, as this would lead to a central intensity minimum (“hole”) and may cause nonadiabatic losses in the dark-state coherence. Therefore, in the present work we restrict the control fields to vortex-free configurations, while the OAM-induced torque originates solely from the probe beams. The extension of the model to include spatially structured control fields will be an interesting direction for future work.

The DT atomic system can be realized experimentally in cold $^{87}\mathrm{Rb}$, where the atomic population is first prepared in a single Zeeman sublevel.
This can be accomplished by applying strong $\sigma^{+}$ laser fields
driving the transitions from the ground hyperfine states $\lvert F=1\rangle$
and $\lvert F=2\rangle$ to the excited states $\lvert F'=2\rangle$
and $\lvert F''=2\rangle$, corresponding to the $\lvert5S_{1/2}\rangle$,
$\lvert5P_{1/2}\rangle$, and $\lvert5P_{3/2}\rangle$ manifolds,
respectively. Under these pumping conditions, the Zeeman sublevel $\lvert F=2,m=2\rangle$
will be the only dark ground state, and optical pumping transfers
the entire population into this state. Two probe fields drive the
transitions $\lvert F=2\rangle\to\lvert F'=2\rangle$ and $\lvert F=2\rangle\to\lvert F''=2\rangle$
with $\sigma^{-}$ polarization. Since all other ground-state Zeeman
sublevels remain unpopulated, the only relevant probe transitions are
$\lvert F=2,m=2\rangle\to\lvert F'=2,m=1\rangle$ and $\lvert F=2,m=2\rangle\to\lvert F''=2,m=1\rangle$. 
This produces DT coupling configuration shown in Fig.~1, with
no significant contribution from other states or transitions. 

\begin{acknowledgments}
This project has received funding from the Research Council of Lithuania
(LMTLT), agreement No.~S-ITP-24-6.
\end{acknowledgments}

\end{document}